\documentclass[sigconf]{acmart}
\usepackage{booktabs} % For formal tables

\setcopyright{none}
\begin{document}

%PPA: Here I use the term preference which strictly refers to comparing items. We would alternatively use the term taste.
% Looking for synonyms but it seems that taste is the best fitting.
% http://www.thesaurus.com/browse/taste
\title{A preference elicitation interface for collecting dense recommender datasets with rich user information}
% \titlenote{Produces the permission block, and
%  copyright information}
\subtitle{Demo}
%\subtitlenote{The full version of the author's guide is available as
%  \texttt{acmart.pdf} document}

\author{Pantelis {P. Analytis}}
%\authornote{Dr.~Trovato insisted his name be first.}
%\orcid{1234-5678-9012}
\affiliation{%
  \institution{Cornell University}
%  \streetaddress{P.O. Box 1212}
%  \city{Ithaca} 
%  \state{NY} 
%  \postcode{43017-6221}
}
\email{pp464@cornell.edu}

\author{Tobias Schnabel}
%\authornote{The secretary disavows any knowledge of this author's actions.}
\affiliation{%
  \institution{Cornell University}
%  \streetaddress{P.O. Box 1212}
%  \city{Dublin} 
%  \state{Ohio} 
%  \postcode{43017-6221}
}
\email{tbs49@cornell.edu}

\author{Stefan Herzog}
% \authornote{This author is the
% one who did all the really hard work.}
\affiliation{%
  \institution{MPI for Human Development}
%  \streetaddress{1 Th{\o}rv{\"a}ld Circle}
%  \city{Hekla} 
%  \country{Iceland}
}
\email{herzog@mpib-berlin.mpg.de}

\author{Daniel Barkoczi}
\affiliation{
  \institution{MPI for Human Development}
%  \streetaddress{P.O. Box 5000}
}
\email{barkoczi@mpib-berlin.mpg.de}

\author{Thorsten Joachims}
\affiliation{%
  \institution{Cornell University}
%  \city{Moffett Field}
%  \state{California} 
%  \postcode{94035}
}
\email{tj@cs.cornell.edu}

%\author{Charles Palmer}
%\affiliation{%
%  \institution{Palmer Research Laboratories}
%  \streetaddress{8600 Datapoint Drive}
%  \city{San Antonio}
%  \state{Texas} 
%  \postcode{78229}}
%\email{cpalmer@prl.com}

% The default list of authors is too long for headers}
% \renewcommand{\shortauthors}{B. Trovato et al.}

\settopmatter{printacmref=false} % Removes citation information below abstract
\pagestyle{plain} % removes running headers

\begin{abstract}

We present an interface that can be leveraged to quickly and effortlessly elicit people's preferences for visual stimuli, such as photographs, visual art and screensavers, along with rich side-information about its users. We plan to employ the new interface to collect dense recommender datasets that will complement existing sparse industry-scale datasets. The new interface and the collected datasets are intended to foster integration of research in recommender systems with research in social and behavioral sciences. For instance, we will use the datasets to assess the diversity of human preferences in different domains of visual experience. Further, using the datasets we will be able to measure crucial psychological effects, such as preference consistency, scale acuity and anchoring biases. Last, we the datasets will facilitate evaluation in counterfactual learning experiments. 

\end{abstract}

%
% The code below should be generated by the tool at
% http://dl.acm.org/ccs.cfm
% Please copy and paste the code instead of the example below. 
%



%\ccsdesc[500]{Computer systems organization~Embedded systems}
%\ccsdesc[300]{Computer systems organization~Redundancy}
%\ccsdesc{Computer systems organization~Robotics}
%\ccsdesc[100]{Networks~Network reliability}

% We no longer use \terms command
%\terms{Theory}


%% Used in some conference proceedings e.g. sigplan and sigchi
% \begin{teaserfigure}
%   \includegraphics[width=\textwidth]{sampleteaser}
%   \caption{This is a teaser}
%   \label{fig:teaser}
% \end{teaserfigure}

\maketitle

%PPA: It would be really cool to find a saying that stresses the importance of data for the development of new theories. 
\vspace{-4mm}
\section{Introduction}

Over the last three decades the recommender systems community has made immense progress in the way we represent, understand and learn people's preferences as a function of previously collected explicit or implicit evaluations. Research in recommender systems has by all means increased the quality of the curated and recommended content in the online world. Several large datasets have been a crucial component of this success, as they have commonly functioned as test-beds on which new theories and algorithms have been compared (Movielens, LastFM and Netflix to name just a few). Most of these datasets, however, are very sparse. They contain thousands items and even the most popular among the items have been evaluated only by a small subset of their users. Given the large fraction of missing ratings, it is challenging to accurately estimate even simple quantities like the average quality of an item, especially since the patterns of missing data are subject to strong selection biases \cite{Pradel/etal/12}. This presents fundamental challenges when evaluating recommendation algorithms on sparse datasets. Further, it becomes an obstacle for scholars in the social and behavioral sciences as workarounds have to be developed for dealing with missing values. 

% PPA: The last sentences above can be strengthened further depending on how much space we have available for it).
% PPA: Here is an entry summarizing existing datasets and reporting on their densities. 

To the best of our knowledge, the only dense collaborative filtering dataset was the outcome of the Jester Interface \cite{goldberg2001eigentaste}. The interface curated 100 jokes of various styles and topics. People utilized a slider to evaluate 5 jokes that were presented to them sequentially. The first evaluations were used to estimate people's preferences and to recommend them the remaining jokes. The users continued to read and evaluate jokes until the pool of 100 items was exhausted. In total, more than 70.000 people have evaluated at least some of the jokes, and more than 14.000 have evaluated all the jokes, resulting to a fully evaluated subset of the dataset. 

% which can be easily used beyond the strict scope of recommender systems research to address crucial questions in cognitive science, such as assessing the potential for learning from others and the inherent variability of preferences in different domains of experience \cite{analytis2015you}.

\begin{figure}[htbp!]
\vspace{-1.5mm}
\includegraphics[width= 1\columnwidth]{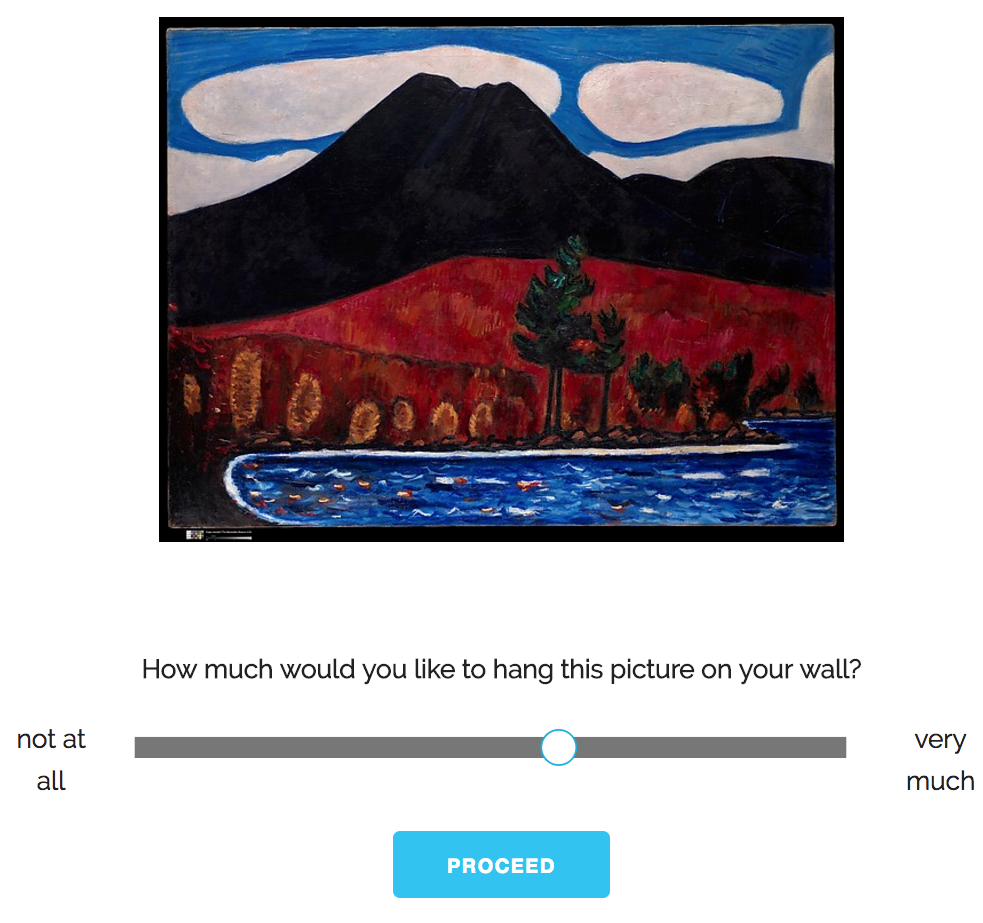}
\vspace{-6mm}
\caption{The design of the preference elicitation interface. We replicate the design of the \emph{Jester} interface, using a continuous bar that people can use to express how much they liked or disliked an item. Participants have to wait for at least 5 seconds before they can proceed to the next item.}
\label{fig:interface}
\vspace{-6mm}
\end{figure}

\section{The interface and data collection}

We plan to collect new datasets in different domains of people's visual experience, ranging from photographs and paintings to designs for screensavers. Our interface replicates the design of the Jester interface, adding new elements that can counteract its limitations. At the outset, people are provided with instructions about how to use the interface. Then, before the presentation of the stimuli we collect demographic information about the users. To reduce possible order effects, the visual stimuli are presented in random order. As in Jester, users are asked to evaluate items using a slider bar; they can move the marker of the slide bar to the left to indicate that they did not like the item, or to the right to indicate that they liked it. We implement a continuous scale, which allows a fine-grained evaluation of the presented items. Finally, to limit anchoring bias, the slide bar is initially semi-transparent and the colors become vivid only when the user has clicked on it.\footnote{The interface can be accessed at \url{http://abc-webstudy.mpib-berlin.mpg.de/recstrgs/study_simulator.php}. Both the code for the interface and the collected data will be publicly available.} 

%{\bf TJ: No sure about the influence of semi-transparency. More important that slider initially has no knob? Would be good if slider was inactive until 6s are over.}

\begin{figure}[htbp!]
 \includegraphics[trim=0 300 0 0,clip,width= \columnwidth]{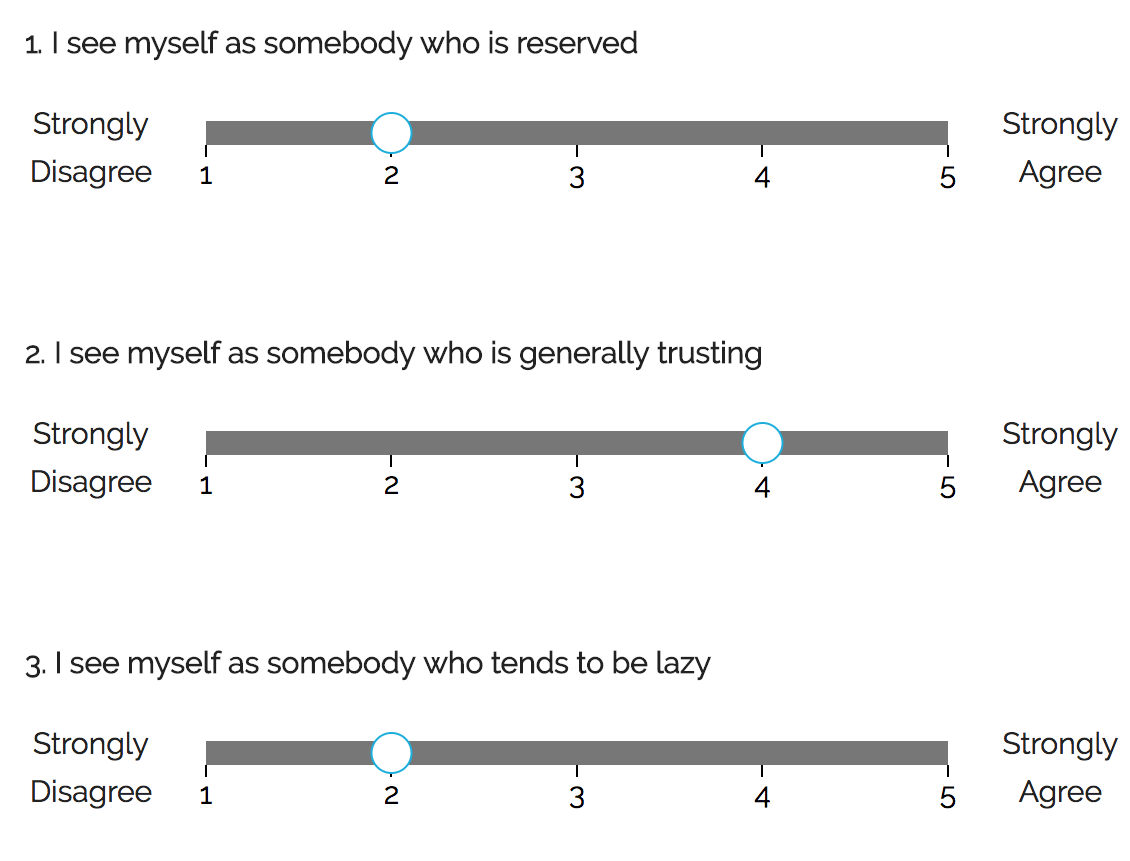}
\caption{At the end of the evaluation phase we collect additional information about the users. We invited the users to complete a questionnaire about their expertist in the visual arts and a 10-question version of the big-five questionnaire.}
\label{fig:questionnaire}
\end{figure}

Once all the items have been evaluated, we collect further psychologically relevant information about the users. Numerous studies have shown that side information can substantially improve estimates of people's preference and it complements first hand evaluations \cite{park2009pairwise}.
%Yet is not always easy to collect it. 
% PPA: In case we find space we can squeeze a reference above. 
In the first experiments we will deploy the visual-art expertise questionnaire developed by Chatterjee et al. \cite{chatterjee2010assessment} to gauge people's familiarity with the visual arts and a succinct version of the big-five questionnaire to quickly assess the people's personalities   \cite{rammstedt2007measuring} (see Figure \ref{fig:questionnaire}). It takes about 20 minutes to complete the current version of the interface, including the instructions, questionnaires and evaluation phase. 

 We intend to conduct the first experiments at Amazon's Mechanical Turk labor market.  Several studies have shown that for effortless tasks the results produced on mTurk are comparable to laboratory studies \cite{paolacci2010running}.
 The visual stimuli used in this interface evoke immediate aesthetic judgments, and thus can quickly be transformed to evaluations. Eventually, we intend to develop a data visualization tool that will reward people who complete the study with information about their preference profiles and how they relate to those of other individuals. Thus, we intend to create an inherently motivating interface using as a reward the informational value generated by the collected data. In this way, we will reduce the cost of data collection, but also introduce basic ideas behind collaborative filtering and recommender systems to the wider public.

\vspace{-3mm}
\section{Potential applications}

We envisage several new applications for the developed datasets. Here we foreshadow a few of these potential uses, keeping in mind that the community that will have access to the produced datasets will certainly come up with more. First, they will facilitate cross-fertilization with the cognitive and behavioral sciences. For instance, social and cognitive psychologists have extensively studied simple strategies for inference and estimation where different features are used to predict an objective truth. The new datasets will open the way to study strategies for social preference learning in domains where no objective truth exists \cite{analytis2015you}. Also, we can manipulate the design of the interface to study relevant behavioral effects, such as to study the consistency of evaluations or to investigate the effect of the granularity of the evaluation scale on the predictions. To sum up, the datasets will allow us to better understand preference diversity and its implications for different recommender systems algorithms as well as for psychological social learning strategies.

Moreover, we believe that the new datasets can fuel existing streams of research in recommender systems and machine learning. For instance, dealing with selection-biases and with data missing not at random is a growing research stream in recommender systems and machine learning \cite{schnabel2016unbiased}. To evaluate algorithms tuned to deal with such problems, we can impose selection biases ex-ante and remove data from the dense dataset accordingly. This set-up could complement existing sparse datasets for learning, with the difference that selection biases can be controlled and varied in order to test robustness. Moving on to the broader class of counterfactual simulations, dense datasets greatly simplify evaluation since they can serve as ground-truth when conducting simulations \cite{salganik2006experimental}. 

% From the perspective of a collaborative system engineer, we can see the opinions of different individuals as different draws from a multi-armed bandit. The mean, and the standard deviation of the empirical distribution correspond closely to the mean of a bandit-arm and the uncertainty describing it. However, in real life we rarely see the evaluations of all the users as the algorithms curating the content favor content that has already accumulated positive evaluations. 

%\end{document}  % This is where a 'short' article might terminate

%\begin{acks}
%We would like to thank Ed Vesser, Ryu Uchiyama, Aman Agarwal, Ashudeep Singh, Angela Zhou and Caroline Chang for their valuable comments on the design of the interface.
%\end{acks}

\bibliographystyle{plain}

\end{document}